\documentclass[conference]{IEEEtran}
\IEEEoverridecommandlockouts
\usepackage{cite}
\usepackage{amsmath,amssymb,amsfonts}
\usepackage{algorithmic}
\usepackage{graphicx}
\usepackage{textcomp}
\usepackage{xcolor}
\def\BibTeX{{\rm B\kern-.05em{\sc i\kern-.025em b}\kern-.08em
    T\kern-.1667em\lower.7ex\hbox{E}\kern-.125emX}}

\begin{document}

\title{EmoTech: A Multi-modal Speech Emotion Recognition Using Multi-source Low-level Information with Hybrid Recurrent Network}

\author{
    \IEEEauthorblockN{Shamin Bin Habib Avro, Taieba Taher, Nursadul Mamun}
    \IEEEauthorblockA{
       Robust Speech Processing Laboratory (RSPL)\\ Department of Electronics and Telecommunication Engineering\\ Chittagong University of Engineering and Technology, Chittagong
       \\ ovrohabib@gmail.com, taieba.athay@cuet.ac.bd, nursad.mamun@cuet.ac.bd
       }
}







\maketitle

\begin{abstract}
Emotion recognition is a critical task in human-computer interaction, enabling more intuitive and responsive systems. This study presents a multimodal emotion recognition system that combines low-level information from audio and text, leveraging both Convolutional Neural Networks (CNNs) and Bidirectional Long Short-Term Memory Networks (BiLSTMs). The proposed system consists of two parallel networks: an Audio Block and a Text Block. Mel Frequency Cepstral Coefficients (MFCCs) are extracted and processed by a BiLSTM network and a 2D convolutional network to capture low-level intrinsic and extrinsic features from speech. Simultaneously, a combined BiLSTM-CNN network extracts the low-level sequential nature of text from word embeddings corresponding to the available audio. This low-level information from both speech and text is then concatenated and processed by several fully connected layers to classify the speech emotion. Experimental results demonstrate that the proposed EmoTech accurately recognizes emotions from combined audio and text inputs, achieving an overall accuracy of 84\%. This solution outperforms previously proposed approaches for the same dataset and modalities.

\end{abstract}

\begin{IEEEkeywords}
Speech Emotion Recognition, Multimodal, BiLSTM, CNN, Text, MFCC
\end{IEEEkeywords}

\section{Introduction}

Emotion recognition from speech has gained significant attention due to its wide range of applications in human-computer interaction, mental health monitoring, and customer service \cite{zhou2018emotional,madanian2022automatic,petrushin1999emotion}. The ability to accurately recognize emotions in speech can enhance the effectiveness of automated systems, making interactions more natural and responsive to users' emotional states. Traditional methods for emotion recognition have primarily focused on using either audio or text features independently. However, leveraging both modalities simultaneously can potentially improve the performance of Speech Emotion Recognition (SER) systems by capturing complementary information from both audio and text data.

Several studies have been proposed in recent years to incorporate audio and text to improve emotion recognition \cite{cho2018deep,sarker2023text, tripathi2018multi,yenigalla2018speech}. For instance, Lee et al. proposed an SER system using text and audio features with a logical "OR" function at the decision level to combine acoustic and linguistic information \cite{lee2002combining}. Later, Jin et al. suggested combining auditory and linguistic information and training them with an SVM classifier to identify emotion categories \cite{jin2015speech}. Recent advancements in deep learning have also been utilized for emotion classification. Griol et al. trained three datasets using machine learning classifiers to categorize user emotions based on spoken utterances \cite{griol2019combining}. Yoon et al. employed RNN networks for both audio and text, while Atmaja et al. proposed LSTM and dense network architectures for emotion classification \cite{yoon2018multimodal, atmaja2019speech}.

This study proposes EmoTech, a multi-modal architecture for speech emotion recognition that combines both audio and text features at a low-feature level. In the proposed model, separate processing blocks are employed for audio and text inputs, which are subsequently concatenated and fed into a classification block to predict the emotion of the speech. The key contributions of this study include:

\begin{enumerate}
\item A multimodal architecture that effectively integrates audio and text modalities for emotion recognition.
\item The use of BiLSTM and CNN layers in both audio and text blocks to capture temporal dependencies and local features.
\item A comprehensive evaluation of the proposed model on standard speech emotion datasets, demonstrating its effectiveness compared to existing methods.

\end{enumerate}
The paper is organized as follows: Section 2 describes the individual parameters of the proposed network used in this research. Section 3 presents the experimental results, followed by the conclusion in Section 4.

\section{Methodology}
This section presents the overall architecture of the proposed EmoTech for SER. It also describes the hyperparameter tuning of the network, along with the training and testing procedures, and the relevant dataset used for this study.

\subsection{Dataset}

\begin{figure}[htbp]
\centerline{\includegraphics[width=0.5\textwidth, keepaspectratio]{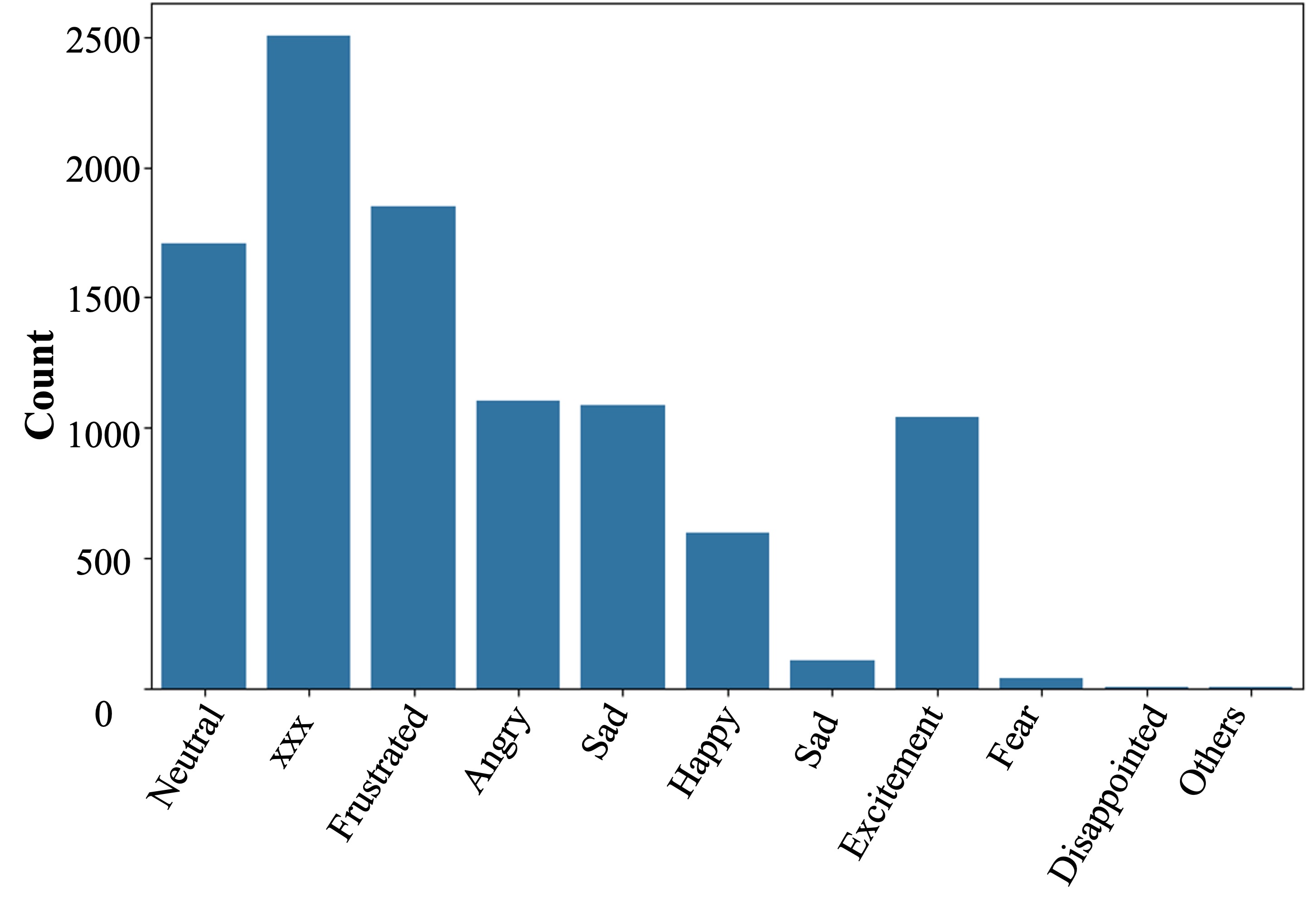}}
\caption{The statistics of ten classes in IEMOCAP Dataset before augmentation}
\label{fig: Unbalanced_dataset}
\end{figure}

This study utilized the Interactive Emotional Dyadic Motion Capture (IEMOCAP)\cite{busso2008iemocap} dataset to train and evaluate the proposed model. The IEMOCAP dataset, a widely used benchmark in speech emotion recognition research, comprises scripted and spontaneous acts of ten emotions. As shown in Fig. ~\ref{fig: Unbalanced_dataset}, the dataset is unbalanced across these ten emotion categories, with some emotions having significantly fewer samples. To address this imbalance, we focused on the five dominant emotion categories: anger, sadness, happiness, neutrality, and excitement.

\begin{figure}[htbp]
\centerline{\includegraphics[width=0.5\textwidth, keepaspectratio]{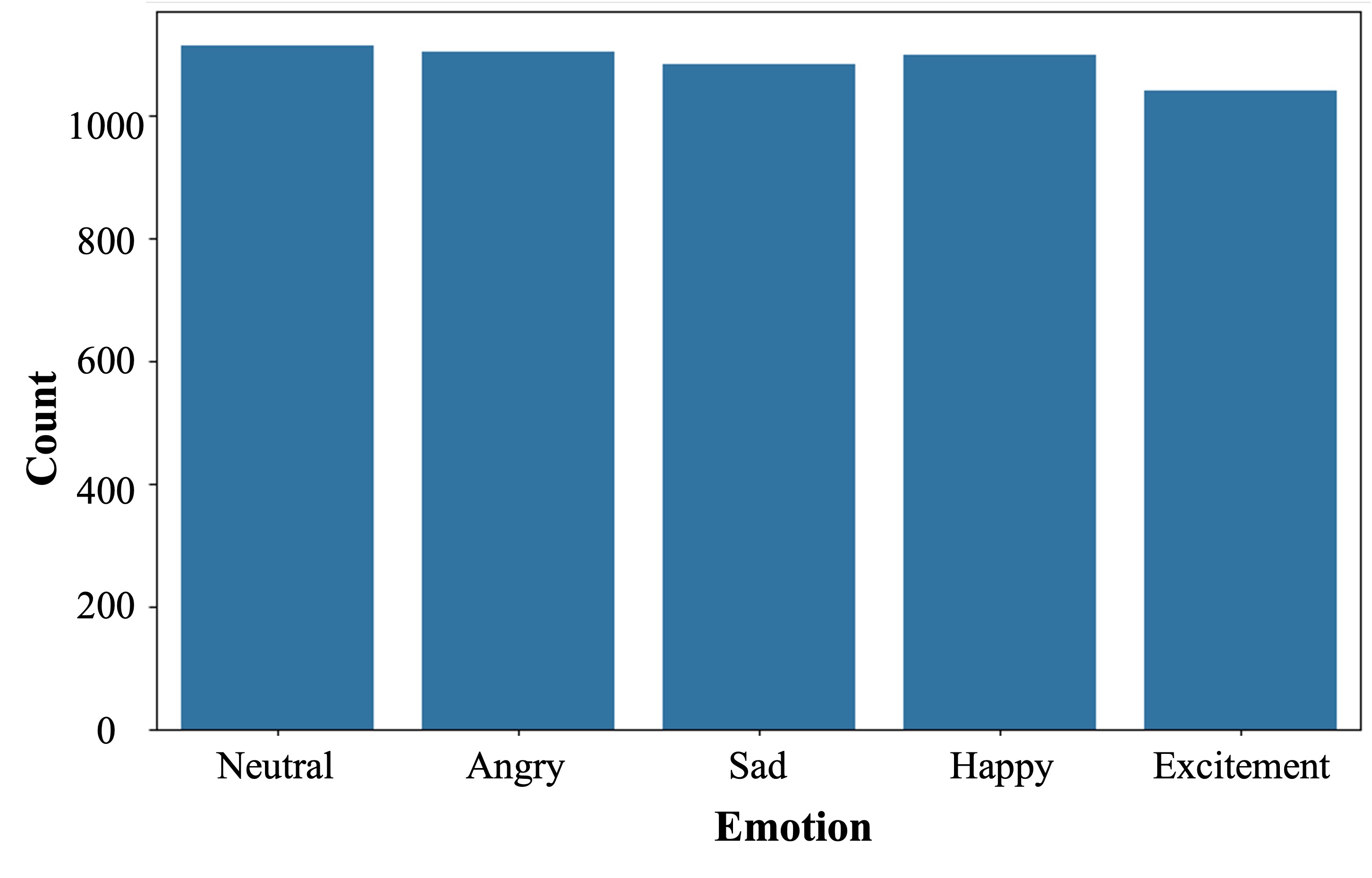}}
\caption{The statistics of five dominated classes in IEMOCAP Dataset after augmentation}
\label{fig: Balanced_dataset}
\end{figure}

Both audio recordings and text transcriptions from the dataset were used. To increase the number of samples in the less dominant classes and achieve a more balanced dataset, various data augmentation techniques were applied. For audio data, techniques such as time stretching, pitch shifting, noise addition, time shifting, and volume adjustment were used. For text data, augmentation techniques included synonym replacement, random insertion, random deletion, and random swapping. The statistics of the total 5,633 data samples after augmentation are represented in Fig. ~\ref{fig: Balanced_dataset}.

\subsection{Network Architecture}
\begin{figure*}[!t]
\centerline{\includegraphics[width=0.75\textwidth, keepaspectratio]{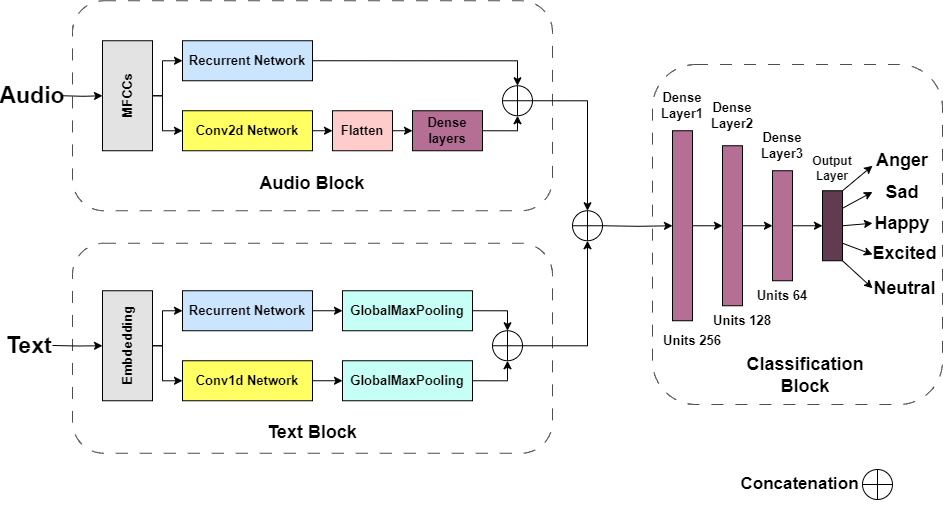}}
\caption{Basic block diagram of the proposed EmoTech Architecture}
\label{fig: Overall_Block_Diagram}
\end{figure*}

Figure ~\ref{fig: Overall_Block_Diagram} illustrates the basic block diagram of the proposed EmoTech, designed for SER. The network comprises three main blocks: 1) Audio Block, 2) Text Block, and 3) Classification Block. The input audio data is first transformed into Mel Frequency Cepstral Coefficients (MFCCs) and processed through a recurrent network to capture temporal dependencies and a 2D convolutional network to extract spatial features. In parallel, the text data is converted into embeddings and processed through a recurrent network with global max pooling to capture sequential information. Additionally, the embeddings pass through a 1D convolutional network with global max pooling to convert them into low-dimensional feature maps. The outputs from the audio and text blocks are then concatenated, passed through several dense layers with decreasing units, and finally classified into one of the emotion categories: anger, sad, happy, excited, or neutral.

\subsubsection{Audio Block}
\begin{figure}[htbp]
\centerline{\includegraphics[width=0.5\textwidth, keepaspectratio]{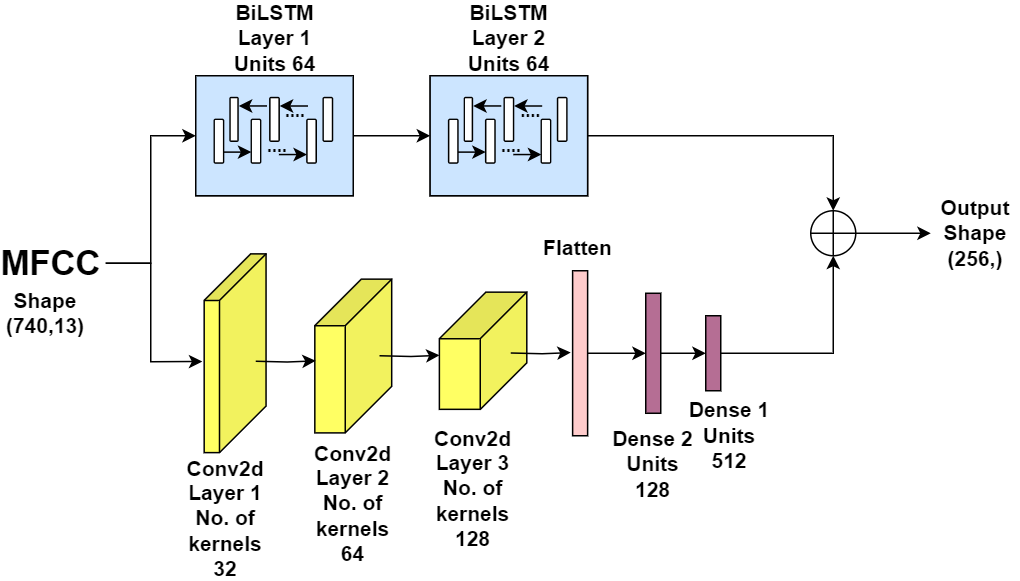}}
\caption{A detail block diagram of audio signal processing in EmoTech}
\label{fig: Audio_Block}
\end{figure}

The detailed block diagram of the audio block is presented in Fig. ~\ref{fig: Audio_Block}. In EmoTech, MFCCs are utilized as the input for the audio block. Initially, sentence-level utterances are resampled to a sampling rate of 16 kHz, and silenced parts are removed using a threshold value of 20 dB. The number of cepstral coefficients used is 13, with the FFT window length set to 2048 (12.8 ms) and the hop length to 512. The 'Hann' window function is applied, and zero padding ensures that all MFCCs have a consistent shape. Consequently, the final shape of the MFCCs for each utterance is (740, 13), where 740 represents the number of time steps and 13 represents the number of cepstral coefficients.

The MFCCs with a shape of (740, 13) are then passed as input to the BiLSTM network. The BiLSTM network comprises two layers, each with 64 hidden units and employing the 'tanh' activation function. The output of the BiLSTM network results in a shape of (128,).

 Simultaneously, the same MFCCs are used as input for the Conv2D network. This network consists of three Conv2D layers, each followed by batch normalization and a MaxPooling layer. The convolution layers have 32, 64, and 128 kernels, respectively, with each kernel having a shape of (3, 3). Each MaxPooling layer utilizes a window shape of (2, 2).

Following the flatten operation in the Conv2D network, a tensor with a shape of (11776,) is generated and used as input for the dense network within the audio block. This dense network's output has a shape of (128,). The dense network comprises two layers, with the first dense layer having 512 hidden units and the second dense layer having 128 hidden units. A dropout layer with a dropout rate of 0.2 is placed between the two dense layers.

Finally, the outputs of the BiLSTM network (with an output shape of (128,)) and the dense network (also with an output shape of (128,)) are concatenated. This concatenation results in the final output of the audio block having a shape of (256,).

\subsubsection{Text Block}

\begin{figure}[htbp]
\centerline{\includegraphics[width=0.5\textwidth, keepaspectratio]{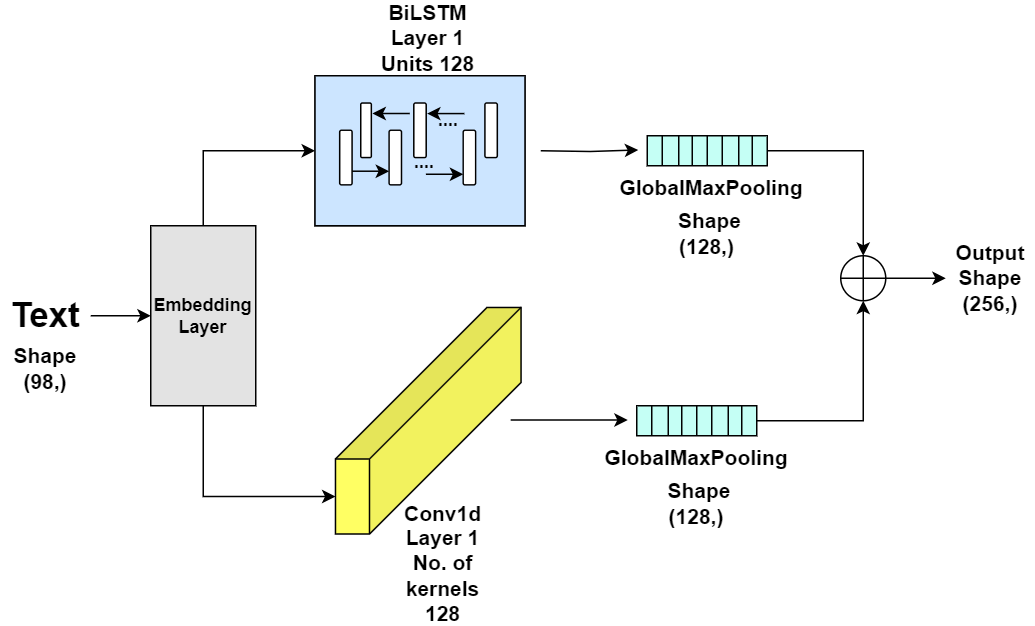}}
\caption{A detail block diagram of text data processing in EmoTech}
\label{fig: Text_Block}
\end{figure}

As shown in Fig. ~\ref{fig: Text_Block}, the corresponding transcriptions of the audio signals are utilized as the text input in this study. Each text sentence is tokenized and zero-padded to ensure uniform length. These tokenized sentences are then processed through an embedding layer. The maximum length of a sentence is set to 98 words, and the vocabulary consists of 2843 words. The integer-encoded sentences, having a shape of (98,), are used as input for the embedding layer. The number of embedding dimensions is 200, resulting in an output shape of (98, 200).

The output of the embedding layer is then used as input for two parallel networks: a BiLSTM network and a Conv1D network. The BiLSTM network is configured with 64 hidden units. The output of the BiLSTM layer is passed into a GlobalMaxPooling layer, resulting in an output shape of (128,). Simultaneously, the output of the embedding layer is used as input for the Conv1D network, which uses 128 kernels with a kernel size of 5, and applies the 'ReLU' activation function for each convolution layer. Following the Conv1D layer, a GlobalMaxPooling layer is utilized, also resulting in an output shape of (128,).

The final output of the text block is achieved by concatenating the outputs of both the Conv1D and BiLSTM networks, resulting in a shape of (256,).

\subsubsection{Classification Block}

\begin{figure}[htbp]
\centerline{\includegraphics[width=0.5\textwidth, keepaspectratio]{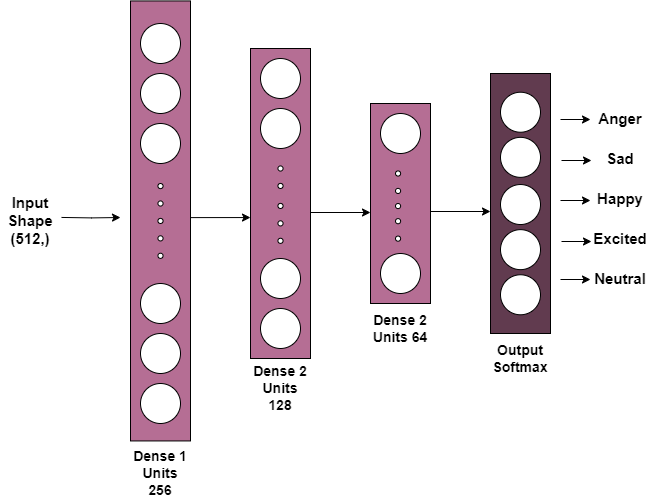}}
\caption{A detail block diagram of classification processing in EmoTech}
\label{fig: Classification_Block}
\end{figure}

This section of EmoTech combines the low-level features from the audio and text blocks, as shown in Fig. ~\ref{fig: Classification_Block}. The classification block consists of three dense layers with unit sizes of 256, 128, and 64, respectively. A dropout layer with a rate of 0.2 follows the first two dense layers to prevent overfitting. The final output layer is configured with 5 units, corresponding to the five emotion classes. Each of the dense layers is activated using the 'ReLU' function, except for the output layer, which uses the 'Softmax' activation function.

\subsection{Hyperparameters and Model Training}

Table 1 represents the different hyperparameters used to train the proposed EmoTech network. The proposed model is trained on 5,633 data samples using 5-fold cross-validation to ensure robust performance evaluation and prevent overfitting.

A total of 30 epochs is specified, with a batch size set to 32. The optimization process is carried out using the Adam optimizer. The model's loss is calculated through the Categorical Cross Entropy function. An initial learning rate of 0.001 is defined, while the minimum learning rate is adjusted to 0.000001 during training.

To further optimize training, EarlyStopping is employed to halt training when the validation loss ceases to improve, and ReduceLROnPlateau is used to reduce the learning rate when the validation performance plateaus, thereby enhancing the model's convergence efficiency.

The total number of parameters in the model is 7,295,821, indicating a complex architecture designed to capture intricate patterns in the data. The training process is conducted on Google Colab, utilizing a T4 GPU as the accelerator to expedite computations and leverage the parallel processing capabilities of the GPU. This setup allows for efficient handling of the model's extensive parameter space and accelerates the training process significantly.

\section{Results and Discussions}
This section presents the simulated results of the proposed EmoTech algorithm, evaluated through objective metrics including precision, recall, and F1-score. The performance of EmoTech is compared with three existing algorithms to provide a benchmark. Additionally, the impact of different emotions on the accuracy of the algorithm is analyzed and discussed.

\subsection{Effect of data augmentation on EmoTech}

Table \ref{table: Augmentation_result} presents the impact of data augmentation on minority classes and feature modality in terms of classification accuracy before and after augmentation. The classification accuracy is shown for different modalities. Generally, the overall accuracy of the model is higher when EmoTech uses combined speech and text for classification rather than a single feature modality.

When combining time-domain audio features and text-embedding features, EmoTech demonstrated the best performance, regardless of augmentation. Moreover, the classification accuracy is higher after augmentation than before augmentation, irrespective of the feature modality used. Therefore, the combination of augmented speech and text is utilized as features for the proposed network and for further evaluation.

\begin{table}
\caption{The impact of data augmentation on accuracy}
\begin{center}
\resizebox{\columnwidth}{!}{ 
\tiny
\begin{tabular}{ccc}
\hline\hline
\textbf{Feature} & \textbf{Augmentation} & \textbf{Accuracy} \\
\hline\hline
\\[-1em] 
Speech & No & 0.7022 \\[0.5em]
Speech & Yes & 0.7184  \\[0.5em]
Text & No & 0.7133  \\[0.5em]
Text & Yes & 0.7423  \\[0.5em]
Speech + Text & No & 0.8105  \\[0.5em]
Speech + Text & Yes & 0.8352  \\[0.5em]
\hline
\end{tabular}
}
\label{table: Augmentation_result}
\end{center}
   \vspace{-10pt}
\end{table}

\subsection{Performance evaluation for different classes }

Table \ref{table: Overall_result} presents the accuracy of individual emotion classes for SER. The accuracy metrics are provided for five emotion classes: anger, sadness, happiness, neutrality (denoted as "neu"), and excitement. These metrics include precision, recall, F1-score, and overall accuracy.
\begin{table}
\caption{Different classification metrics for individual classes}
\label{tab:my_label}

\begin{center}
\resizebox{\columnwidth}{!}{ 
\begin{tabular}{ccccc}
\hline\hline
\textbf{Class} & \textbf{Precision} & \textbf{Recall} & \textbf{F1-score} & \textbf{Accuracy}\\
\hline\hline
\\[-1em] 
Anger & 0.9641 & 0.9729 & 0.9685 & 0.9728 \\[0.5em]
Excited & 0.9083 & 0.9252 & 0.9167 & 0.9252 \\[0.5em]
Happy & 0.9224  & 0.9456  & 0.9339 & 0.9456 \\[0.5em]
Neutral & 0.8660 & 0.8153 & 0.8399 & 0.8153 \\[0.5em]
Sad & 0.9654 & 0.9696 & 0.9675  & 0.9695 \\[0.5em]

\hline
\end{tabular}
}
\label{table: Overall_result}
\end{center}
   \vspace{-20pt}
\end{table}

In general, the accuracy is notably high for the "angry" and "sad" emotions, demonstrating robust performance in both precision and recall. Conversely, the "neu" (neutrality) emotion tends to have lower accuracy compared to the others. Specifically, the highest accuracy score observed is 97.28\% for the "angry" emotion.

\begin{figure}[htbp]
\centerline{\includegraphics[width=0.45\textwidth, keepaspectratio]{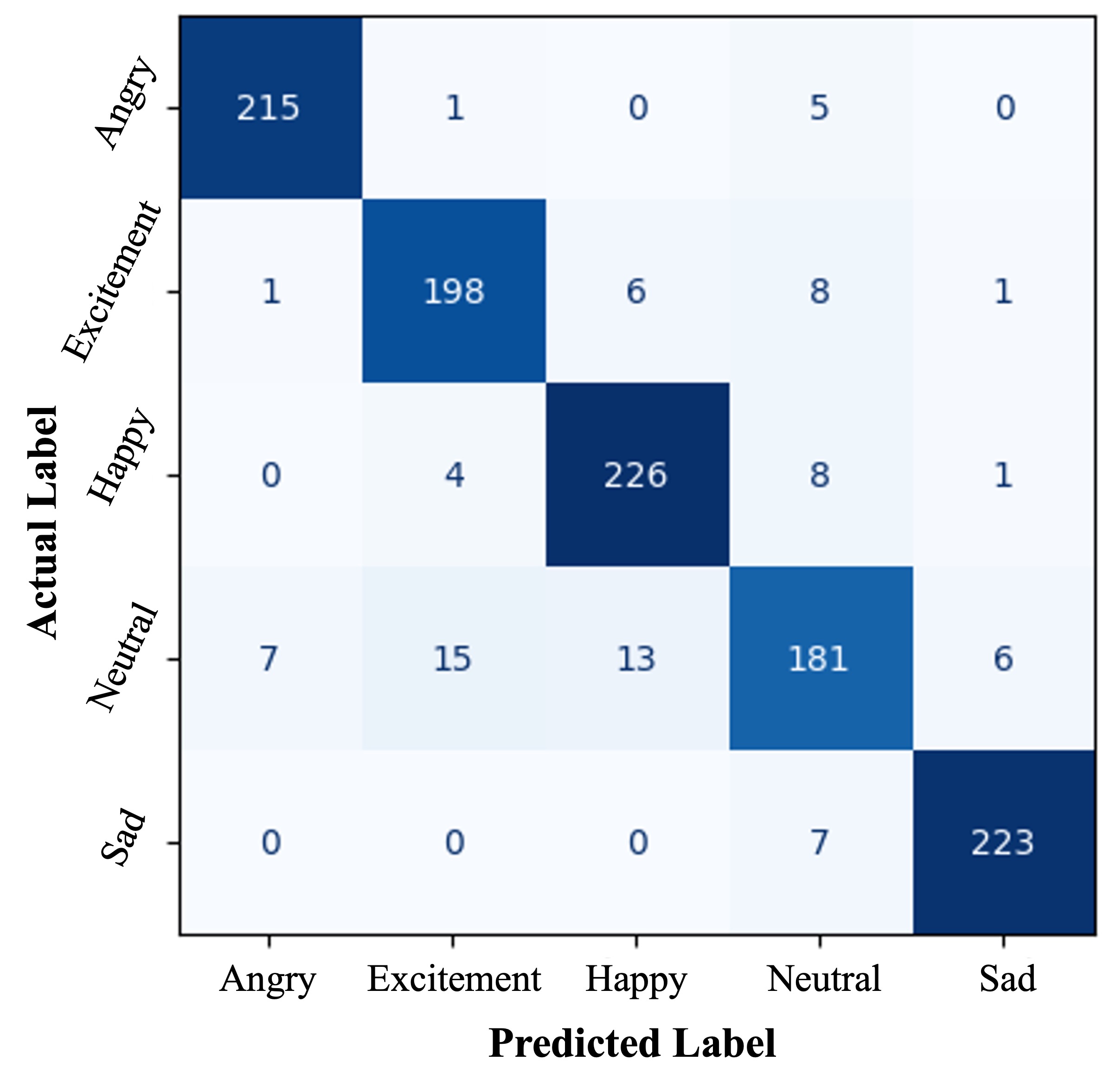}}
\caption{Confusion Matrix in classifying five emotion using EmoTech architecture.}
\label{fig: Confusion_Matrix}
\end{figure}

The confusion matrix for individual emotions is shown in Fig. \ref{fig: Confusion_Matrix}. The results indicate that the neutral emotion is frequently misclassified, particularly confused with excitement and happiness.

\subsection{Comparison with existing networks }

To analyze the performance of the proposed network over existing methods, three different models are evaluated with the IEMOCAP dataset. The results are evaluated in terms of overall accuracy and presented in Table \ref{table: Comparison_table}. In general, the accuracy is high when a hybrid model is used to capture the detailed information from speech and text. The result shows that the score is high when emotions are classified using the proposed EmoTech network.

\begin{table}
\caption{Comparison of different algorithms in SER}
\begin{center}
\resizebox{\columnwidth}{!}{ 
\begin{tabular}{ccc}
\hline\hline
\textbf{Model} & \textbf{Feature} & \textbf{Accuracy(\%)} \\
\hline\hline
\\[-1em] 
Yoon\cite{yoon2018multimodal} & Speech+Text & 71.80  \\[0.5em]
Yenigalla\cite{yenigalla2018speech} & Speech+Phoneme & 73.90  \\[0.5em]
Atmaja\cite{atmaja2019speech} & Speech+Text & 75.40  \\[0.5em]

\textbf{EmoTech} & \textbf{Speech+Text} & \textbf{83.52}  \\[0.5em]
\hline
\end{tabular}
}
\label{table: Comparison_table}
\end{center}
\end{table}

\section{Conclusion}
This study introduces a multi-modal architecture for SER that effectively integrates audio and text features to enhance classification performance. By incorporating BiLSTM and CNN layers within dedicated audio and text blocks, the model successfully captures both temporal dependencies and local features, thereby improving its ability to detect nuanced emotional expressions. Evaluation on the IEMOCAP dataset, a widely accepted benchmark in emotion recognition, illustrates that the proposed model surpasses traditional single-modal approaches and achieves competitive accuracy comparable to state-of-the-art methods. Future research directions include extending the modalities to include audio, text, and video inputs, thereby broadening the scope of emotion recognition applications.

\bibliography{Manuscript}

\bibliographystyle{plain}

\end{document}